\documentclass[aps,twocolumn,a4paper,showpacs,superscriptaddress]{revtex4}
\usepackage{tensor}
\usepackage{graphicx}
\usepackage{amsmath}
\usepackage{amssymb}
\usepackage{enumerate}
\usepackage{subfigure}
\usepackage{tabularx}
\usepackage[colorlinks=true, pdfstartview=FitV, linkcolor=blue, citecolor=red, urlcolor=black, breaklinks=true]{hyperref}

\newcommand{\be}{\begin{equation}}
\newcommand{\ee}{\end{equation}}
\newcommand{\ben}{\begin{eqnarray}}
\newcommand{\een}{\end{eqnarray}}
\newcommand{\bes}{\begin{subequations}}
\newcommand{\ees}{\end{subequations}}
\def\bal#1\eal{\begin{align}#1\end{align}}
\newcommand{\vphi}{\varphi}

\newcommand{\LL}{{\mathcal L}}

\begin{document}
\title{Analytic vortex solutions in generalized models of the Maxwell-Higgs type}
\author{D. Bazeia}\affiliation{Departamento de F\'\i sica, Universidade Federal da Para\'\i ba, 58051-970 Jo\~ao Pessoa, PB, Brazil}
\author{L. Losano}\affiliation{Departamento de F\'\i sica, Universidade Federal da Para\'\i ba, 58051-970 Jo\~ao Pessoa, PB, Brazil}
\author{M.A. Marques}\affiliation{Departamento de F\'\i sica, Universidade Federal da Para\'\i ba, 58051-970 Jo\~ao Pessoa, PB, Brazil}
\author{R. Menezes}\affiliation{Departamento de F\'\i sica, Universidade Federal da Para\'\i ba, 58051-970 Jo\~ao Pessoa, PB, Brazil}\affiliation{Departamento de Ci\^encias Exatas, Universidade Federal da Para\'{\i}ba, 58297-000 Rio Tinto, PB, Brazil}
\begin{abstract}
This work deals with the presence of analytical vortex configurations in generalized models of the Maxwell-Higgs type in the three-dimensional spacetime. We implement a procedure that allows to decouple the first order equations, which we use to solve the model analytically. The approach is exemplified with three distinct models that show the robustness of the construction. In the third model, one finds analytical solutions that exhibit interesting compact vortex behavior. 
\end{abstract}
\date{\today}
\maketitle

\section{Introduction}

Vortices are planar configurations that appear in several areas of nonlinear science. In high energy physics, in particular, vortices first appeared in \cite{NO}, in a model governed by the Maxwell field coupled to a complex scalar field with standard covariant derivative in the presence of a Higgs-like potential. This is the standard Maxwell-Higgs model, and it was later shown to support first-order differential equations that solve the equations of motion \cite{B,VS}. The use of the Maxwell-Higgs model in applications of practical interest can be found, for instance, in Refs.~\cite{B1,B2} and in references therein.

The problem can be studied in $(2,1)$ spacetime dimensions, and it is a known fact that, despite the presence of first order equations that solve the equations of motion, the standard Maxwell-Higgs model is hard to be solved analytically. However, the standard model is composed by the addition of three distinct contributions, the first governing the Maxwell dynamics, the second describing the covariant derivative and the third adding the Higgs-like potential. Thus, if one thinks of generalizing the model with the inclusion of no other field, a good possibility is to add new functions of the scalar fields, modifying the Maxwell dynamics and/or multiplying the covariant derivative. In the generalized model which will be investigated in the current work we consider the case of adding two new functions.

A motivation to study generalized models comes from the context of Cosmology, as studied in Refs.~\cite{kinf,cosm1,cosm2}, for instance. An interesting feature of these models is that the inflation may be driven without the presence of a potential. Following the same lines, defect structures were considered in models with generalized kinematics in Refs.~\cite{babichev1,babichev2}. Over the years, several papers have investigated similar generalizations; see Refs.~\cite{kd1,kd2,kd3,kd4,kd5,kd6,kd7,kd8}. Distinct generalization were considered before in several works, in particular in \cite{L1,L2,L3} to generalize chromodielectric properties of the system, in \cite{G1,G2}, with the aim to simulate Chern-Simons vortices using a generalized Maxwell-Higgs model, and in \cite{ana}, to implement a gauge embedding procedure that produces the dual mapping of the self-dual vector field theory to a Maxwell-Chern-Simons model. It is worth commenting that compact vortices only appear if generalized models are considered \cite{compm,compcs}. The addition of a function of the scalar field multiplying the Maxwell term or the covariant derivative term is also present in models related to holography; see, e.g., \cite{H0,H1,H2} for three distinct possibilities, used to describe an holographic superconductor that admits an analytic treatment near the phase transition \cite{H0}, an holographic insulator model with nonsingular zero temperature infrared geometry \cite{H1}, and the electric charge transport in a strongly coupled quark-gluon plasma \cite{H2}.

The study of analytic vortex solutions in generalized models is also related to another issue of current interest, which concerns the presence of two gauge and two scalar fields, changing the $U(1)$ symmetry to become $U(1)\times U(1)$. The enlarged model is of interest to describe superconducting strings, as proposed in Ref.~\cite{W}; see also Refs~\cite{B1,W1,W2} and references therein for other related issues. In this case, the coupling occurs between the two scalar fields in the Higgs potential. The enlarged model is also of interest to include the so-called {\it hidden} sector, and one possible way to couple the {\it visible} sector to a {\it hidden} one is via the Higgs fields, as in the so-called Higgs portal \cite{D1,D2}, in a way similar to the coupling used in \cite{W,W1,W2}. But we can also make the two gauge fields interact in the way suggested for instance in \cite{DM1,DM2,DM3,DM4,DM5}, with the two gauge field strengths coupled to each other. However, if one works with generalized models, there are other possible ways to couple a {\it visible} $U(1)$ gauge model with another {\it hidden} $U(1)$ gauge model, for instance using a function of the scalar field of the hidden sector to modify the permeability and/or the dynamics of the scalar field of the visible sector, among other possibilities. Thus, we believe that the presence of analytic vortex solutions in generalized models of the form to be studied in the current work will contribute to shed light on how the {\it visible} sector may interact with a {\it hidden} sector.

We also remind the investigations \cite{ana1,kd8,ana3,compm,FO2}, in which the authors study several issues, in particular the possibility of finding analytic solutions for generalized Maxwell-Higgs models. Since the presence of analytic solutions is welcome, and since the previous studies only offer specific investigations, in this work we focus mainly on the possibility of constructing a general procedure, from which one can describe generalized Maxwell-Higgs models that support analytic vortex configurations. Moreover, the presence of analytic solutions may be of use to shed light in the compactification of vortices studied in Refs.~\cite{compm,compcs}, whose calculations are mostly numerical.

The nonlinearities of the field equations make the problem very hard, but we have gotten inspiration from the recent works \cite{compcs,FO2}, mainly the last one, where one implements a first order formalism that help us to describe generalized Maxwell-Higgs, Chern-Simons-Higgs, and Maxwell-Chern-Simons-Higgs models \cite{FO2}. In particular, we also study the possibility of describing analytical vortices that attain the compact behavior \cite{compcs}. We implement the investigation in the next Sec.~\ref{sec:2}, describing the procedure in a detailed way, including the decoupling of the first order equations and explaining the construction of the generalized model. We add another section, Sec.~\ref{sec:3}, which is dedicated to illustrate the procedure with three distinct examples. We conclude in Sec.~\ref{sec:4}, where we add some comments and possible new investigations that deserve further consideration.

\section{The general procedure}\label{sec:2}

We consider vortices in $(2,1)$ spacetime dimensions governed by the generalized Lagrangian density
\be\label{lmodel}
\LL =- \frac{1}{4}P(|\vphi|)F_{\mu\nu}F^{\mu\nu}+ K(|\vphi|)|D_{\mu}\vphi|^2  - V(|\vphi|),
\ee
where $F_{\mu\nu}=\partial_{\mu}A_\nu-\partial_{\nu}A_\mu$ is the electromagnetic tensor, $D_{\mu}=\partial_{\mu}+ieA_{\mu}$ stands for the covariant derivative, and $V(|\vphi|)$ is the potential for the scalar field. Also, $P(|\vphi|)$ is added to describe generalized magnetic permeability and $K(|\vphi|)$ is used to modify the covariant derivative contribution. We take the Minkowski metric $\eta_{\mu\nu}=(1,-1,-1)$ and use natural units, with $\hbar=c=1$. 

The possibility to enlarge the model to comply with the $U(1)\times U(1)$ symmetry will be briefly discussed at the very end of the work. In the current study we investigate the $U(1)$ model \eqref{lmodel}, focusing attention on the construction of analytic solutions. The equations of motion associated to the Lagrangian density \eqref{lmodel} are
\bes\label{geom}
\bal
 D_\mu (K D^\mu\vphi)&= \!\frac{\vphi}{2|\vphi|}\!\left(\!K_{|\vphi|}|D_{\mu}\vphi|^2 \!- \frac{1}{4}P_{|\vphi|}F_{\mu\nu}F^{\mu\nu}\! - V_{|\vphi|}\right), \\ \label{meqs}
 \partial_\mu \left(P F^{\mu\nu} \right) &= J^\nu,
\eal
\ees
where the current is $J_\mu = ieK(\bar{\vphi}D_\mu \vphi-\vphi\overline{D_\mu\vphi})$. Here we are using $K_{|\varphi|}=dK/d|\varphi|$, etc. 
We also have that the energy-momentum tensor has the form
\begin{equation}\label{emt}
T_{\mu\nu}=P F_{\mu\lambda}\tensor{F}{^{\lambda}_{\nu}}+ K\left({\overline{D_{\mu}{\vphi}}}{D_{\nu}{\vphi}}+{\overline{D_{\nu}{\vphi}}}{D_{\mu}{\vphi}}\right)-g_{\mu\nu}{\LL}.
\end{equation}
One can set $\nu=0$ in Eq.~\eqref{meqs} to see that the Gauss' law is solved by taking $A_0=0$, for static field configurations. This implies that the vortex is electrically neutral. 

To search for vortexlike solutions, we consider static configurations and the usual ansatz
\bes\label{ansatz}
\bal
\vphi &= g(r)e^{in\theta},\\
\vec{A} &= -{\frac{\hat{\theta}}{er}(a(r)-n)},
\eal
\ees
where $n$ is an integer number that stands for the vorticity. The functions $a(r)$ and $g(r)$ obey the boundary conditions
\bes\label{bc1}
\bal
g(0) &= 0 \quad \text{and}\quad a(0)=n, \\
g(\infty) &= v \quad \text{and} \quad a(\infty)=0.
\eal
\ees
where $v$ is a parameter used to control spontaneous symmetry breaking of the potential. Considering the ansatz given by Eqs.~\eqref{ansatz}, the magnetic field becomes $B=-a^\prime/(er)$, where the prime stands for the derivative with respect to $r$. By using this, one can show that the magnetic flux is quantized
\be
\begin{split}
\Phi &=2\pi\int_0^\infty rdr B \\
     &=\frac{2\pi}{e}n.
\end{split}
\ee
The equations of motion \eqref{geom} with the ansatz \eqref{ansatz} assume the form
\bes\label{eomansatz}
\begin{align}\label{eomg}
&\frac{1}{r} \left(r K g^\prime\right)^\prime -\frac{ K a^2g}{r^2} = \nonumber
\\
& \frac12 \left(K_{|\vphi|}\left({g^\prime}^2 +\frac{a^2g^2}{r^2}\right) + P_{|\vphi|}\frac{{a^\prime}^2}{2e^2r^2} + V_{|\vphi|}\right), \\ \label{eoma}
& r\left(P\frac{a^\prime}{r} \right)^\prime - 2e^2 K ag^2 = 0.
\end{align}
\ees
The energy density can be calculated from the energy-momentum tensor 
\eqref{emt}. Considering the ansatz \eqref{ansatz}, we can write
\bes
\bal\label{rhoans}
T_{00} &= K(g) \left({g^\prime}^2 +\frac{a^2g^2}{r^2}\right) + P(g) \frac{{a^\prime}^2}{2e^2r^2} + V(g), \\
T_{12} &= K(g) \left( {g^\prime}^2 - \frac{a^2g^2}{r^2} \right) \sin(2\theta), \\ 
T_{11} &= P(g) \frac{{a^\prime}^2}{2e^2r^2} + K(g) \bigg({g^\prime}^2(2\cos^2\theta-1) \nonumber\\
&\hspace{4mm} +\frac{a^2g^2}{r^2}(2\sin^2\theta -1)\bigg) -V(g), \\ 
T_{22} &= P(g) \frac{{a^\prime}^2}{e^2r^2} + K(g) \bigg({g^\prime}^2(2\sin^2\theta-1) \nonumber\\
&\hspace{4mm}+\frac{a^2g^2}{r^2}(2\cos^2\theta-1) \bigg) -V(g).
\eal
\ees

We now follow Ref.~\cite{FO2} to get to a first-order formalism to study vortices. By setting the stressless condition, $T_{ij}=0$, we obtain the first-order equations
\be\label{fo}
g^\prime = \frac{ag}{r} \quad \text{and}\quad a^\prime = -er \sqrt{2V/P}.
\ee
They must be compatible with the equations of motion, so one is left with the constraint
\be\label{constr}
\frac{d}{dg}\sqrt{2VP} = -2egK(g).
\ee
In the above expression, the functions $P(g)$ and $K(g)$ have to be chosen in a way that leaves room for the spontaneous symmetry breaking of the potential $V(g)$.

The first order equations \eqref{fo} can be used in the energy density \eqref{rhoans} to give
\be\label{rhofo}
\rho = P(g) \frac{{a^\prime}^2}{e^2r^2} + 2K(g) {g^\prime}^2.
\ee
This first order formalism includes the possibility of introducing an auxiliar function $W(a,g)$, defined as \cite{FO2}
\be
W_a = P(g) \frac{a^\prime}{e^2r}, \quad \text{and} \quad W_g = 2K(g) r g^\prime,
\ee
where $W$ must obey the constraint $W_g=2agK(g)$ to be compatible with $g^\prime = ag/r$. For the model \eqref{lmodel}, one uses Eqs.~\eqref{fo} to get that
\be\label{w}
W(a,g)=-\frac{a}{e}\sqrt{2V(g)P(g)}.
\ee
The above function allows to write the energy density in the form
\be
\rho=\frac{1}{r} \frac{dW}{dr}
\ee
such that the total energy becomes
\be\label{energyw}
\begin{split}
E &= 2\pi \left|W\left(a(\infty),g(\infty)\right)-W\left(a(0),g(0)\right)\right|\\
&= 2\pi \left|W\left(0,v\right)-W\left(n,0\right)\right|.
\end{split}
\ee
This is an interesting result of Ref.~\cite{FO2}, and shows that the energy only depends on the boundary values of the functions $a(r)$ and $g(r)$ and can be calculated without knowing their explicit behavior. 

Even though the first order equations \eqref{fo} are simpler than the equations of motion \eqref{eomansatz}, they also couple the gauge and matter fields $a(r)$ and $g(r)$ and, in general, can only be solved numerically. For this reason, let us now focus on the possibility of obtaining analytic vortex solutions in the generalized model. In fact, we develop a method that allows to decouple the first-order equations \eqref{fo} and solve them analytically. We then use $V(|\varphi|)$ and the analytical solutions to get $P(|\varphi|)$ and $K(|\varphi|)$, which are required to define the Lagrangian density. For simplicity, however, we consider dimensionless fields and take $e,v=1$; also, we work with unit vorticity, setting $n=1$. 

We implement the procedure introducing the function $R(g)$ such that
\be\label{dg1}
r\frac{dg}{dr}=R(g).
\ee
For a given $R(g)$, we can solve the above equation and find the solution $g(r)$ obeying the boundary conditions \eqref{bc1}. One uses this into Eqs.~\eqref{fo} to obtain
\be\label{arg}
a(r)=\frac{R(g(r))}{g(r)}.
\ee
We then introduce another function, $M(g)$, defined as $M(g)=-\sqrt{2V/P}$. We can use this and the constraint \eqref{constr} to write
\bes\label{VKP}
\bal
V(g)&=\frac12\,P(g)\,M^2(g),\\
K(g)&=\frac{1}{2g}\frac{d}{dg}\left[P(g)M(g)\right].
\eal
\ees
Now, $M(g)$ can be obtained from Eq.~\eqref{fo} to be
\be\label{M}
M(g)=\frac{R(g)}{q^2(g)}\frac{d}{dg}\left(\frac{R(g)}{g}\right).
\ee
In the above expressions, $q(g)$ is the inverse function of $g(r)$, which is assumed to be known for the $R(g)$ that was chosen in Eq.~\eqref{dg1}. This procedure decouples the first order equations in a manner such that the solutions depend exclusively on the function $R(g)$. 

Since $M(g)$ only depends on $R(g)$ and the inverse function of the solution, $q(g)$, we still have room to suggest the potential and use it to get
$K(g)$ and $P(g)$ from Eqs.~\eqref{VKP}. These functions, even though they do not contribute to modify the solution, they play important role in the definition of the model and in the calculation of the energy density, given by Eq.~\eqref{rhofo}. Thus, in the construction of the model one has to be careful with the behavior of the energy.

It is worth commenting that the above prescription to construct the model, given by Eqs.~\eqref{VKP} and \eqref{M}, only requires the functions
$V(|\vphi|)$, $K(|\vphi|)$ and $P(|\vphi|)$ in the interval $|\vphi|\in[0,1]$, since it is there where the solution exists, as one can see from the boundary conditions, Eq.~\eqref{bc1}. Nevertheless, it is important to use non-negative potentials that present a minimum at $|\vphi|=1$, in order to include spontaneous symmetry breaking and protect the models against instabilities and negative energies.

By using Eqs.~\eqref{dg1} and \eqref{arg}, the energy density \eqref{rhofo} can be written in the form
\be\label{rhog}
\rho = \frac{PR^2}{g^2r^4} \left(R_g-\frac{R}{g}\right)^2 + 2 \frac{KR^2}{r^2},
\ee
which will be also used in the applications to be described below.

The procedure described by Eqs.~\eqref{dg1}-\eqref{rhog} seems to be of current interest, but requires further study,
to see how it works under practical consideration. To check this, in the next section we illustrate how it works under several specific applications, including the possibility to describe compact solutions.

\section{Illustrations}
\label{sec:3}

Let us now consider some explicit examples, which give rise to analytical vortex configurations.  

\subsection{Example 1}

\begin{figure}[htb!]
\centering
\includegraphics[width=4.2cm]{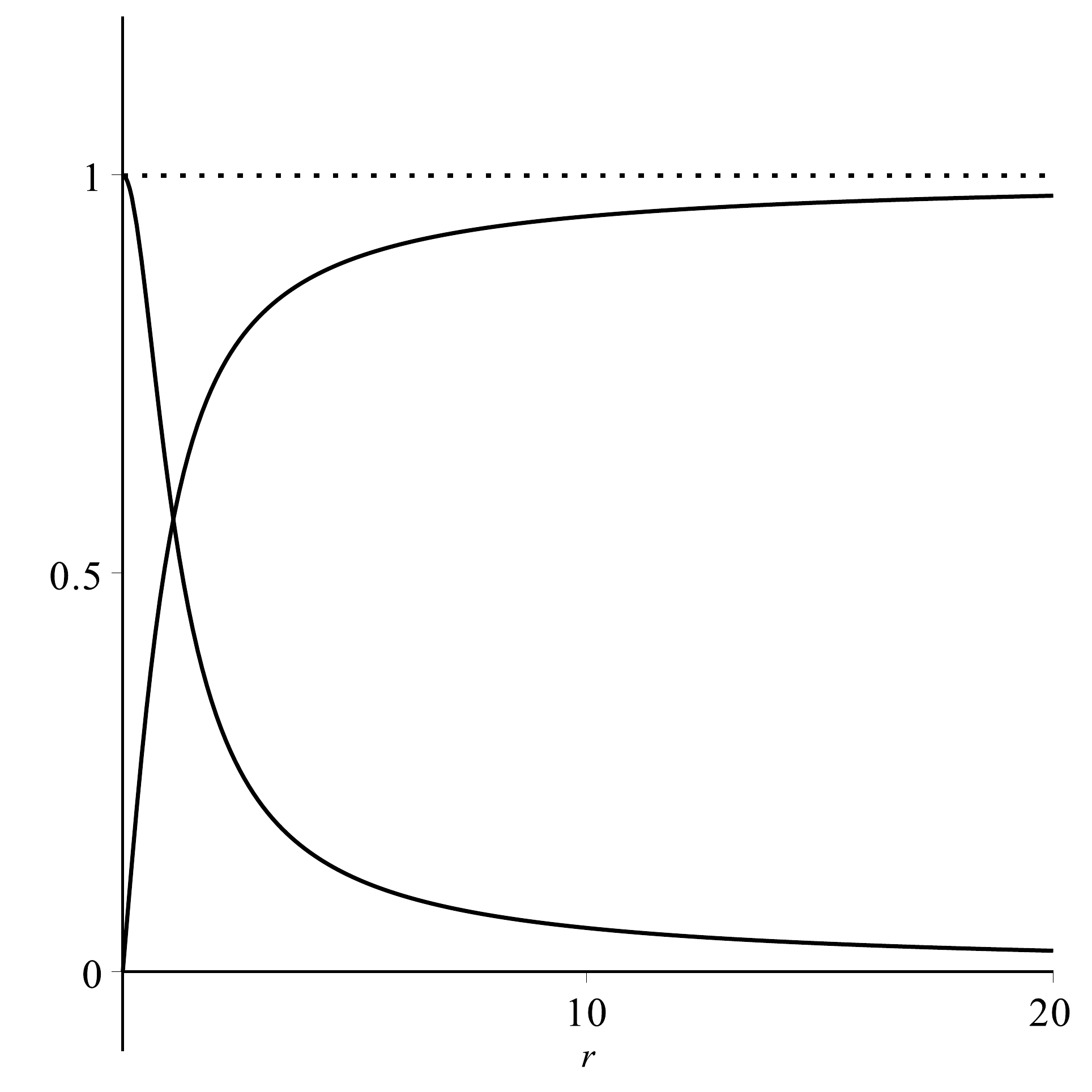}
\includegraphics[width=4.2cm]{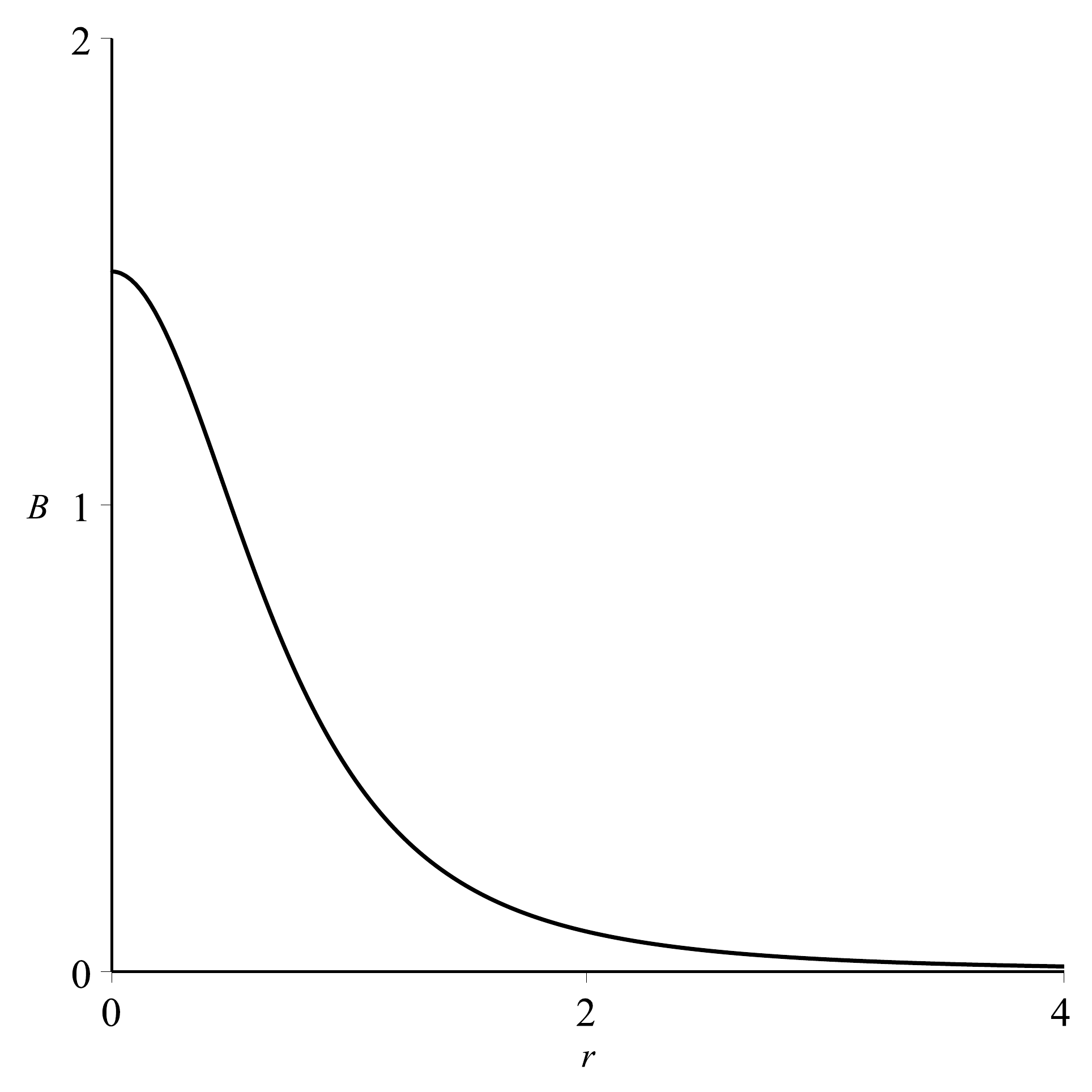}
\caption{In the left panel we depict the solutions $a(r)$ (descending line) and $g(r)$ (ascending line) as in Eqs.~\eqref{ag0}. The right panel shows the magnetic field given by Eq.~\eqref{BR0}.}
\label{figsolb0}
\end{figure}

We first consider the case described by 
\be\label{R0}
R(g)=\frac12\,g\left(1-g^2\right)\left(2-g^2\right).
\ee
If one substitutes the above function in Eqs.~\eqref{dg1} and \eqref{arg}, the following analytical solutions emerge
\bes\label{ag0}
\bal
g(r) &= \sqrt{1-\frac{1}{\sqrt{r^2+1}}}, \\
a(r) &= \frac12\frac{r^2}{\left(r^{2}+1\right)^{3/2}-\left(r^{2}+1\right)}.
\eal
\ees
In this case, the magnetic field is
\be\label{BR0}
B(r)=\frac12\frac{\left(r^{2}-2\right)\sqrt{r^{2}+1}+2}{(r^{2}+1)^2\left(\sqrt{r^{2}+1}-1\right)^2}.
\ee
Notice that both Eqs.~\eqref{ag0} and \eqref{BR0} depends only on the function $R$ given by Eq.~\eqref{R0}. In Fig.~\eqref{figsolb0}, we display the solutions and the magnetic field, to show the general behavior.

\begin{figure}[htb!]
\centering
\includegraphics[width=4.2cm]{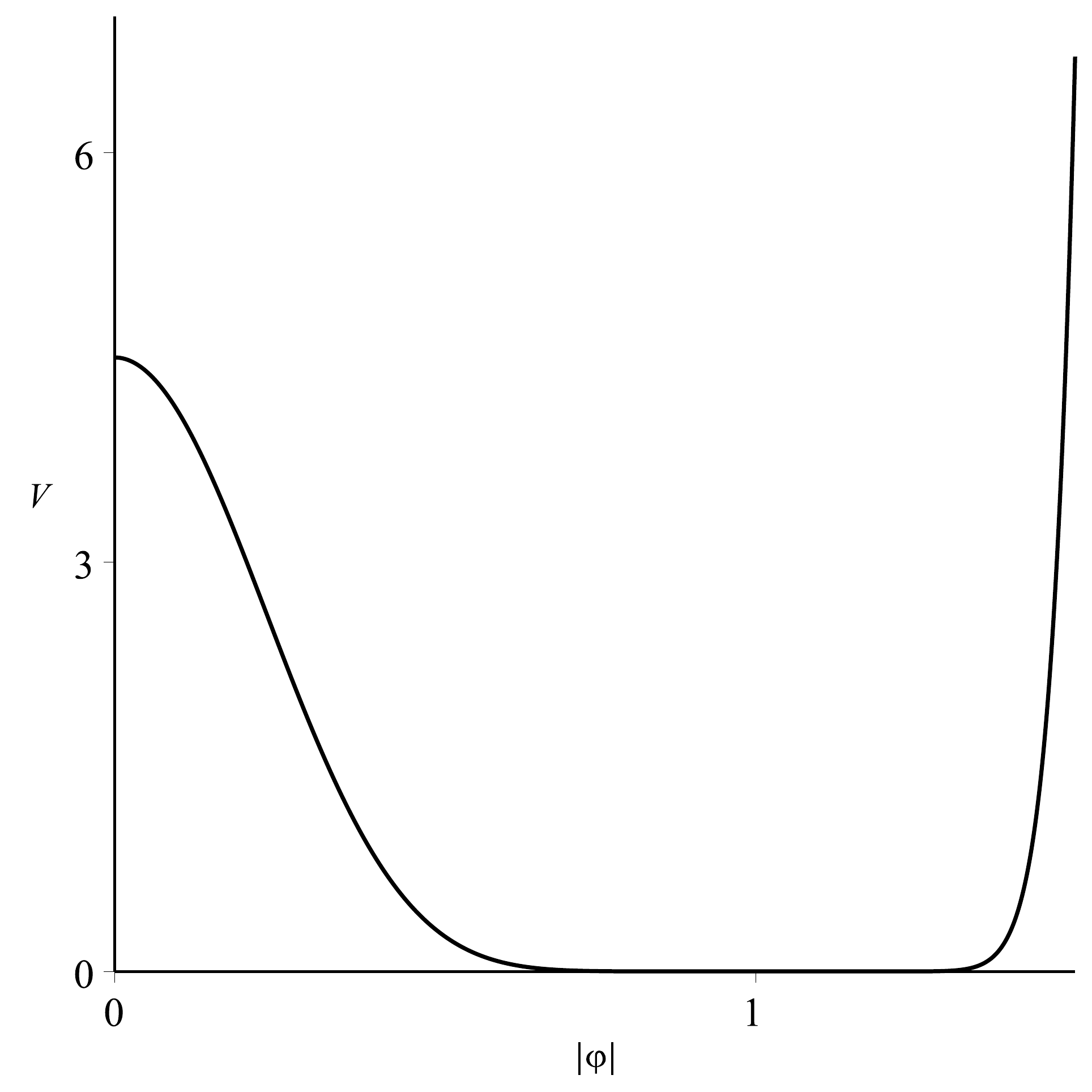}
\includegraphics[width=4.2cm]{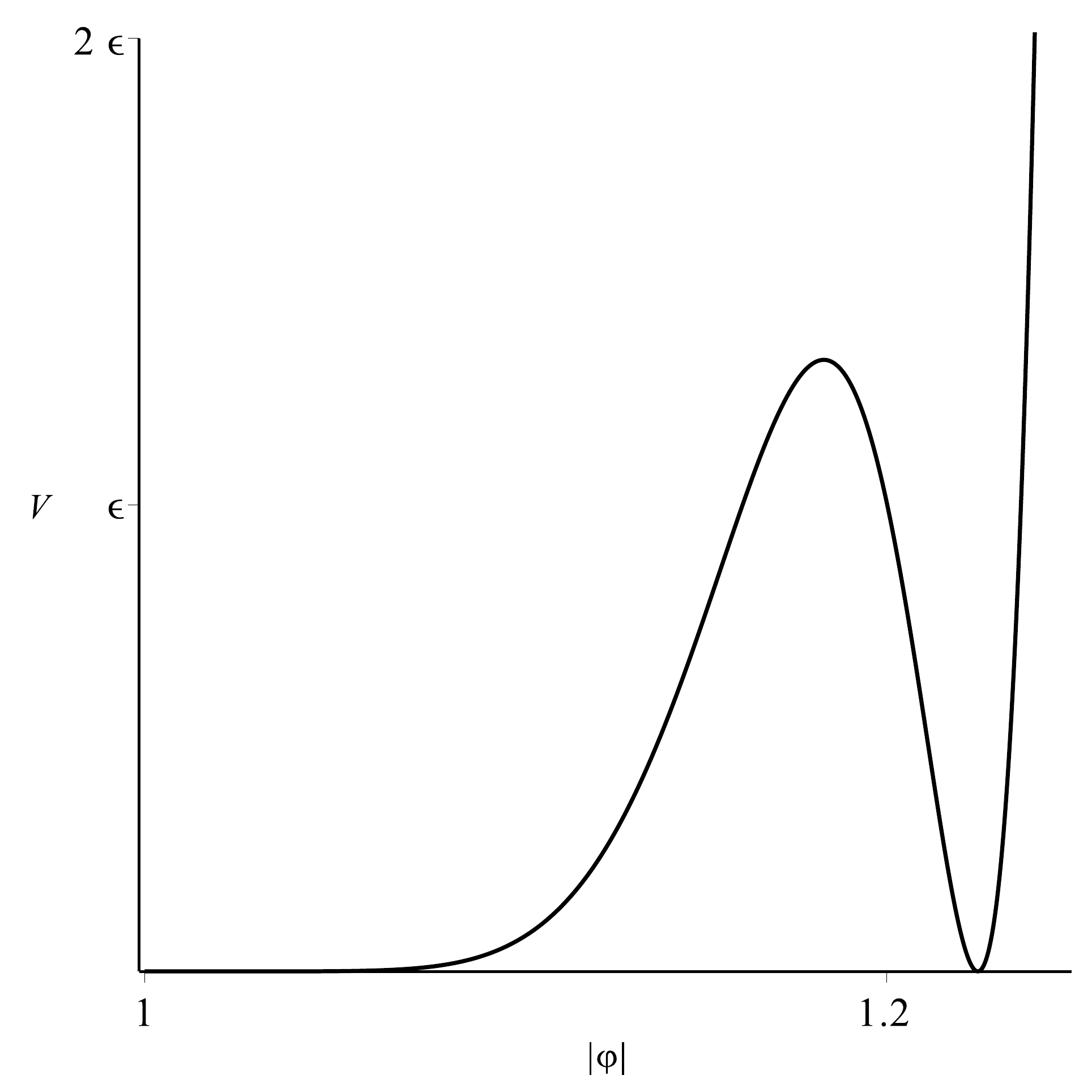}
\caption{The potential of Eq.~\eqref{pot0} in the range $|\vphi|\in[0,1.5]$ (left) and zoomed to the range $|\vphi|\in[1,1.25]$ (right). We have used $\epsilon=10^{-5}$ to adjust the scale of the vertical axis in the right plot.}
\label{figpot0}
\end{figure}

Although the above analytical results are of interest, we have to go further to construct the model. Indeed, we need to find the inverse function of $g(r)$, which we called $q(g)$. Combining it with $R(g)$ in Eq.~\eqref{R0}, we can calculate $M(g)$ by using Eq.~\eqref{M}. The procedure shown in the previous section leads us to
\bal
q(g) &= \frac{g\sqrt{2-g^2}}{1-g^2}, \\
M(g) &= -\frac12 \left(1-g^{2}\right)^3\left(3-2g^2\right).
\eal
To find the functions $K(|\vphi|)$, $P(|\vphi|)$ and $V(|\vphi|)$, we must use Eqs.~\eqref{VKP}. As we discussed before, let us take the potential in the form
\be\label{pot0}
V(|\vphi|) = \frac12 \left(1-|\vphi|^{2}\right)^8\left(3-2|\vphi|^2\right)^2.
\ee
The minima of this potential are at $|\vphi|=1$ and at $|\vphi|=\sqrt{3/2}$. Moreover, it has a local maximum at $|\vphi|=0$, such that $V(0)=9/2$, and another one at $|\vphi| = \sqrt{7/5}$, such that $V\big(\sqrt{7/5}\big) = 128/9765625\approx1.31\times10^{-5}$. In Fig.~\ref{figpot0}, we plot this potential.

\begin{figure}[htb!]
\centering
\includegraphics[width=5cm]{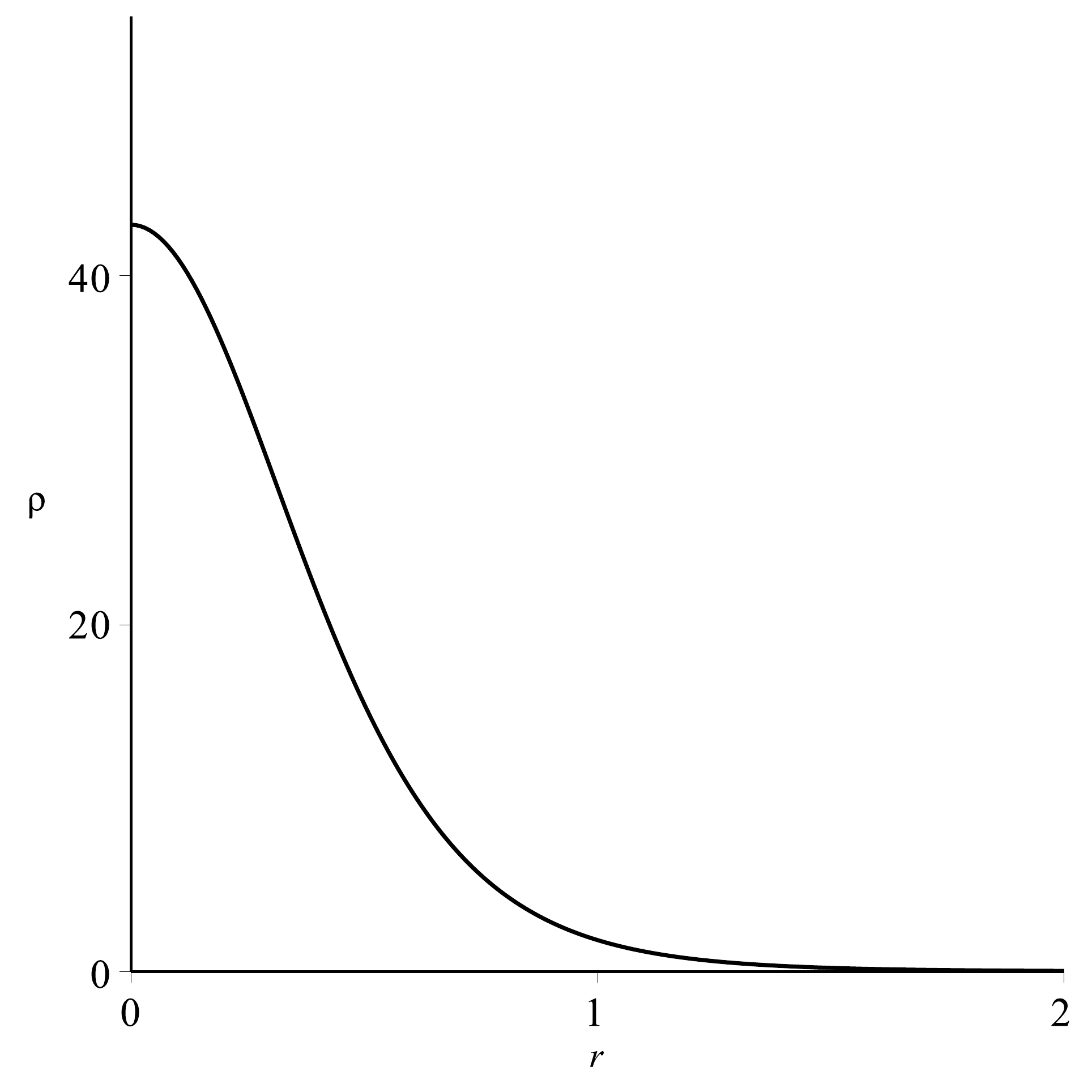}
\caption{The energy density given by Eq.~\eqref{rho0}.}
\label{figrho0}
\end{figure}

Considering the choice of the potential in Eq.~\eqref{pot0}, from Eq.~\eqref{VKP} we can calculate the functions $K(|\vphi|)$ and $P(|\vphi|)$. They are given by
\bal\label{pe}
K(|\vphi|) &= 2 \left(1-|\vphi|^2\right)^4\left|17-12|\vphi|^2\right|,\\
P(|\vphi|) &= 4 \left(1-|\vphi|^2\right)^2.\label{ka}
\eal
Furthermore, one can use Eq.~\eqref{w} to calculate the function $W(a,g)$
\be\label{w0}
W(a,g) = -2 a \left(1-g^2\right)^5\left(3-2g^2\right).
\ee
From Eq.~\eqref{energyw}, it is straightforward to show that the energy is $E=12\pi$.
The energy density can be calculated by using Eq.~\eqref{rhog}. Since the solution exists inside the interval $g\in [0,1]$, we have
\be
\rho(g(r))=\left(1-g(r)^2\right)^8\left(16g(r)^4 - 53g(r)^2 + 43\right).
\ee
By using the solutions \eqref{ag0}, the explicit form of the energy density is
\be\label{rho0}
\rho(r) = \frac{2\left(3r^2+11\right) + 21\sqrt{r^2+1}}{\left(r^2+1\right)^5}.
\ee
In Fig.~\ref{figrho0}, we depict the energy density.
By integrating Eq.~\eqref{rho0}, we get the energy $E=12\pi$, which matches with the value obtained from the auxiliary function $W(a,g)$ in Eq.~\eqref{w0}.

As one can see from the above results, the model is of the form \eqref{lmodel}, with the potential $V(|\varphi|)$ given by Eq.~\eqref{pot0}, and $P(|\varphi|)$ and $K(|\varphi|)$ described by Eqs.~\eqref{pe} and \eqref{ka}, respectively.
\subsection{Example 2}

\begin{figure}[htb!]
\centering
\includegraphics[width=4.2cm]{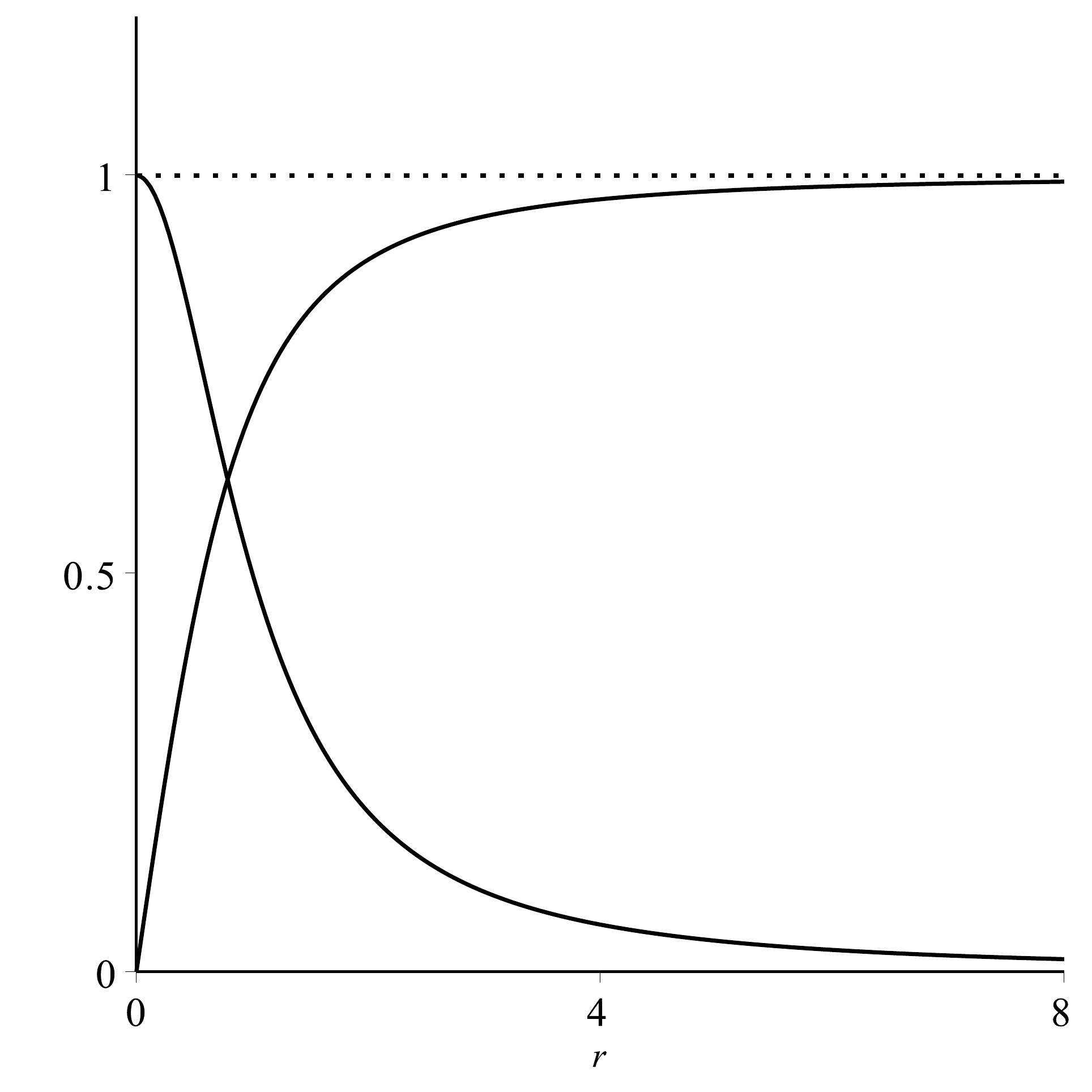}
\includegraphics[width=4.2cm]{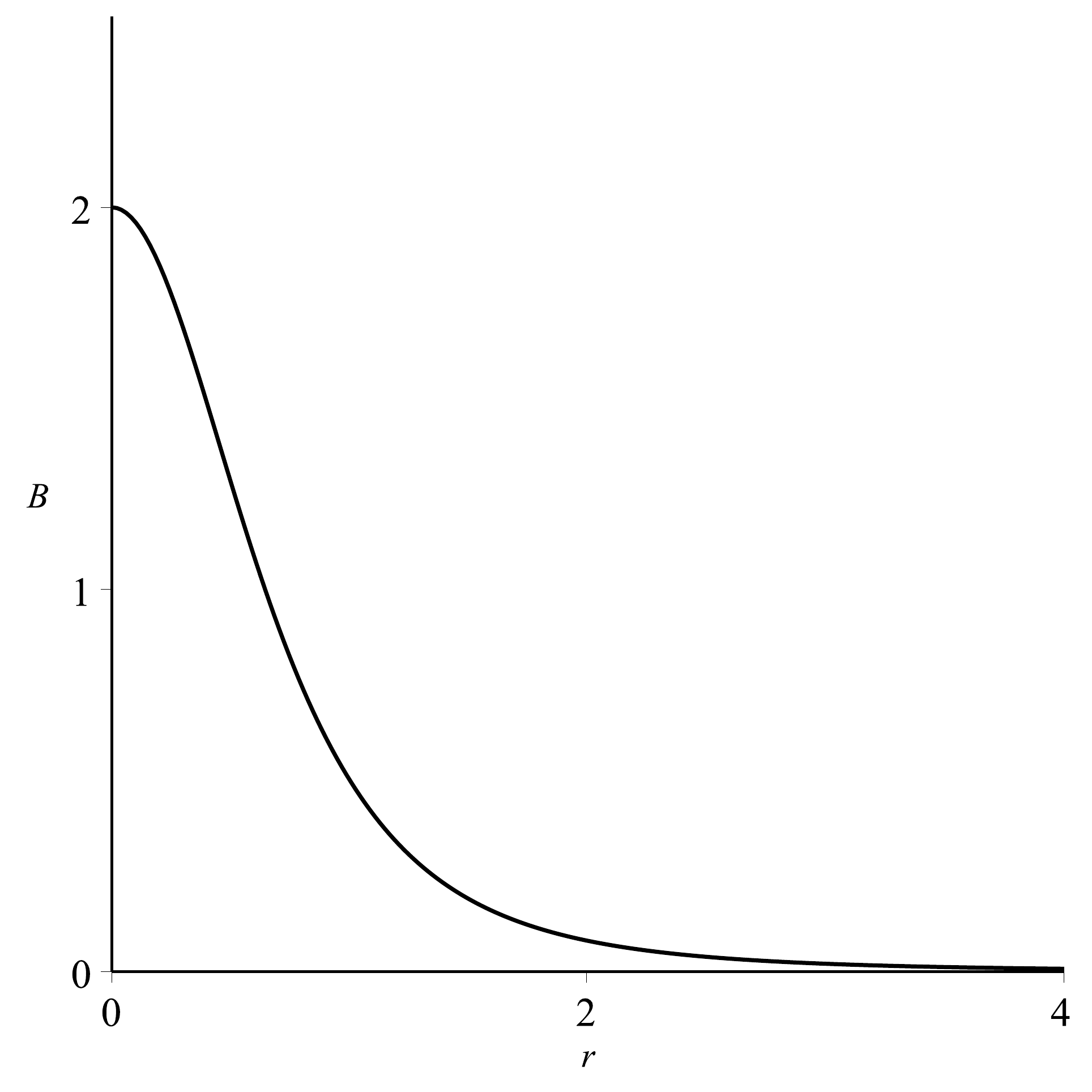}
\caption{In the left panel, we depict the solutions $a(r)$ (descending line) and $g(r)$ (ascending line) as in Eqs.~\eqref{ag1}. The right panel shows the magnetic field given by Eq.~\eqref{BR1}.}
\label{figsolb1}
\end{figure}

\begin{figure}[htb!]
\centering
\includegraphics[width=5cm]{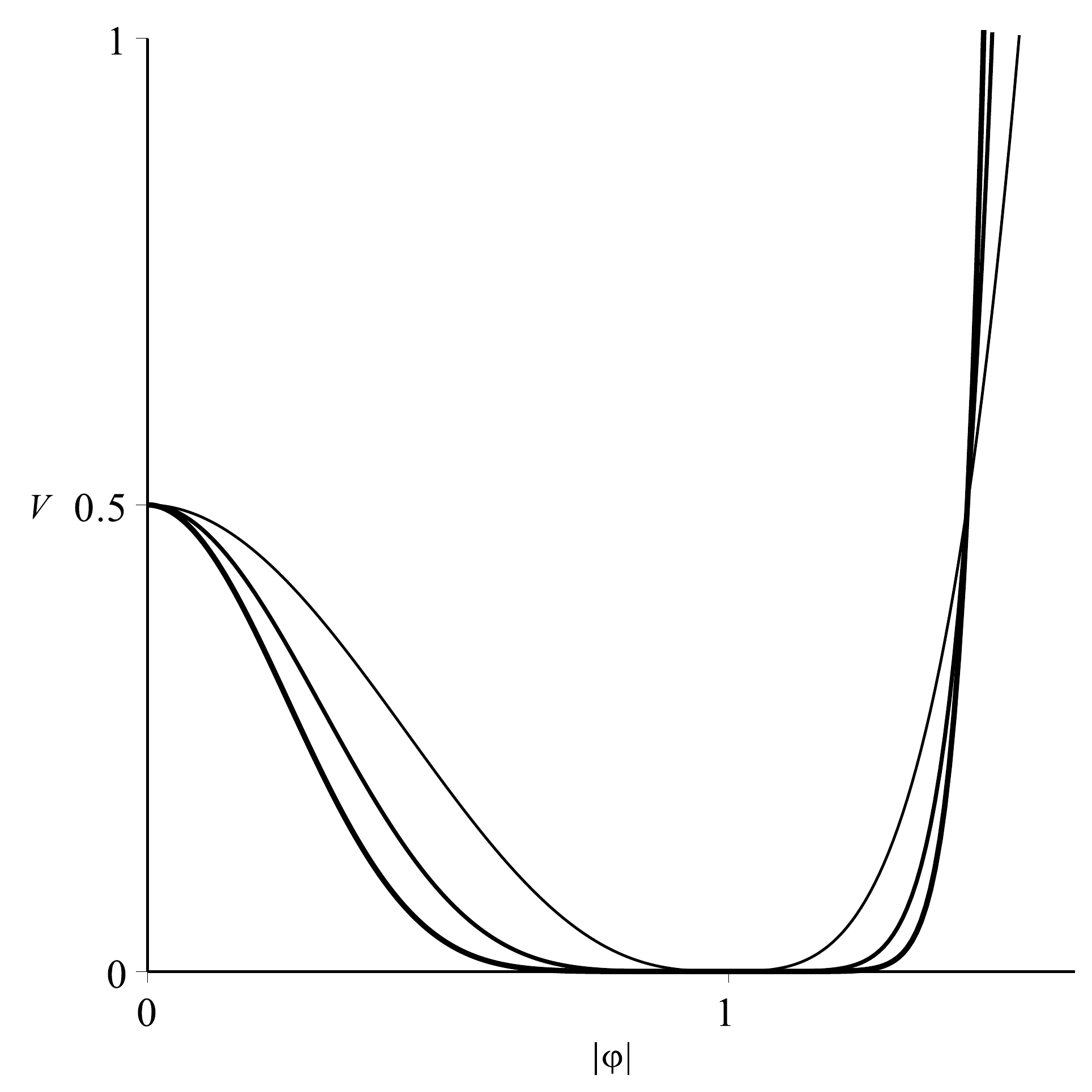}
\caption{The potential in Eq.~\eqref{pot1} for $s=3,6$ and $9$. The thickness of the lines increases with $s$.}
\label{figpot1}
\end{figure}

The second example to be considered is described by 
\be\label{R1}
R(g)=g\left(1-g^2\right).
\ee
By substituting the above function in Eqs.~\eqref{dg1} and \eqref{arg}, one gets the solutions
\be\label{ag1}
g(r) = \frac{r}{\sqrt{1+r^{2}}}\ \quad\text{and}\quad a(r) =\frac{1}{1+r^{2}}.
\ee
These solutions were obtained before in Ref.~\cite{ana1}. Nevertheless, the model was not constructed, with an explicit expression for the Lagrangian in terms of the fields, and this we want to do in the current work. In this case, the magnetic field is
\be\label{BR1}
B(r)=\frac{2}{(1+r^{2})^2}.
\ee
Notice that both Eqs.~\eqref{ag1} and \eqref{BR1} depends only on the function $R$ in Eq.~\eqref{R1}. In Fig.~\eqref{figsolb1}, we plot these functions.
To construct the model, we must find the inverse function of $g(r)$, which we named $q(g)$. Combining it with $R(g)$ in Eq.~\eqref{R1}, we can calculate $M(g)$ by using Eq.~\eqref{M}. Thus, we have
\bal
q(g) = \frac{g}{\sqrt{1-g^{2}}} \quad\text{and}\quad M(g) = -2 \left(1-g^{2}\right)^{2}.
\eal
To find the functions $K(|\vphi|)$, $P(|\vphi|)$ and $V(|\vphi|)$, we must use Eqs.~\eqref{VKP}. As we discussed before, one can take the potential in the form
\be\label{pot1}
V(|\vphi|) = \frac12 \left|1-|\vphi|^2\right|^s,
\ee
where $s$ is a real number such that $s>2$. This potential presents a set of minima at $|\vphi|=1$ and a local maximum at $|\vphi|=0$, such that $V(0)=1/2$. In Fig.~\ref{figpot1}, we plot this potential for $s=3,6$ and $9$. Notice that, as $s$ increases, the concavity of the potential gets wider. This behavior plays no role in the solutions given by Eqs.~\eqref{ag1}.

\begin{figure}[htb!]
\centering
\includegraphics[width=5cm]{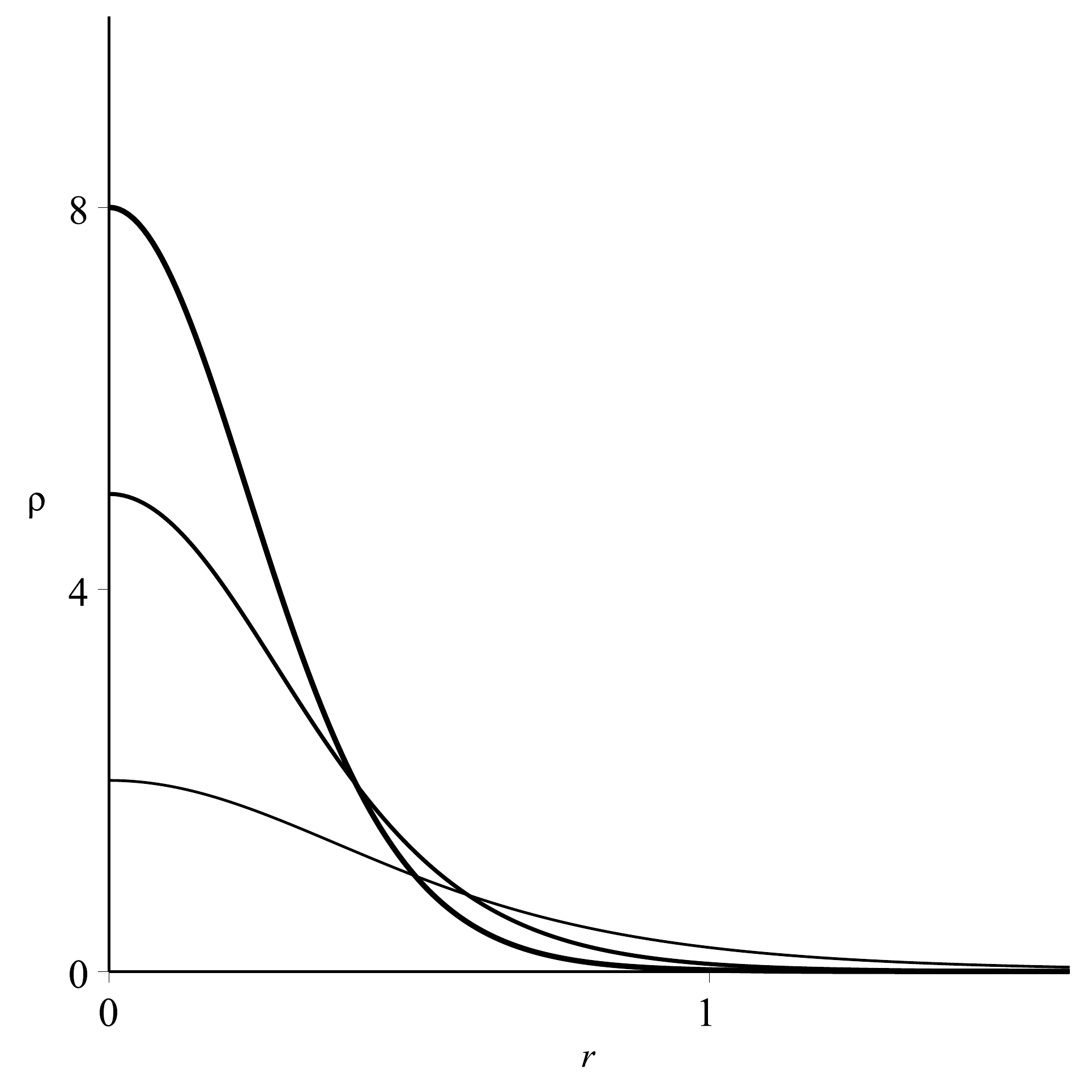}
\caption{The energy density given by Eq.~\eqref{rho1}, depicted for $s=3,6$ and $9$. The thickness of the lines increases with $s$.}
\label{figrho1}
\end{figure}

Considering the choice of the potential in Eq.~\eqref{pot1}, from Eq.~\eqref{VKP} we can calculate the functions $K(|\vphi|)$ and $P(|\vphi|)$. They are given by
\bal\label{ka2}
K(|\vphi|) &= \frac{s-2}{2} \left|1-|\vphi|^2\right|^{s-3},\\
P(|\vphi|) &= \frac{1}{4} \left|1-|\vphi|^2\right|^{s-4}.\label{pe2}
\eal
Furthermore, one can use Eq.~\eqref{w} to calculate the function $W(a,g)$
\be\label{w1}
W(a,g) = -\frac12 a \left(1-g^2\right)^{s-2}.
\ee
From Eq.~\eqref{energyw}, it is straightforward to show that the energy is $E=\pi$.
The energy density can be calculated using Eq.~\eqref{rhog}. Since the solution exists inside the interval $g\in [0,1]$, we have
\be
\rho(g(r))=(s-1)\left(1-g(r)^2\right)^s.
\ee
By using the solutions \eqref{ag1}, the explicit form of the energy density is
\be\label{rho1}
\rho(r) = \frac{s-1}{\left(1+r^2\right)^s}.
\ee
In Fig.~\ref{figrho1}, we plot the energy density for $s=3,6$ and $9$.
By integrating the energy density in Eq.~\eqref{rho1}, we get the energy $E=\pi$, which matches with the value obtained from the auxiliary function $W(a,g)$ in Eq.~\eqref{w1}. 

This model also has the Lagrangian in the form \eqref{lmodel}, with the potential given by \eqref{pot1}, and the functions $K(|\varphi|)$ and
$P(|\varphi|)$ described by \eqref{ka2} and \eqref{pe2}, respectively.

\subsection{Example 3}

\begin{figure}[htb!]
\centering
\includegraphics[width=4.2cm]{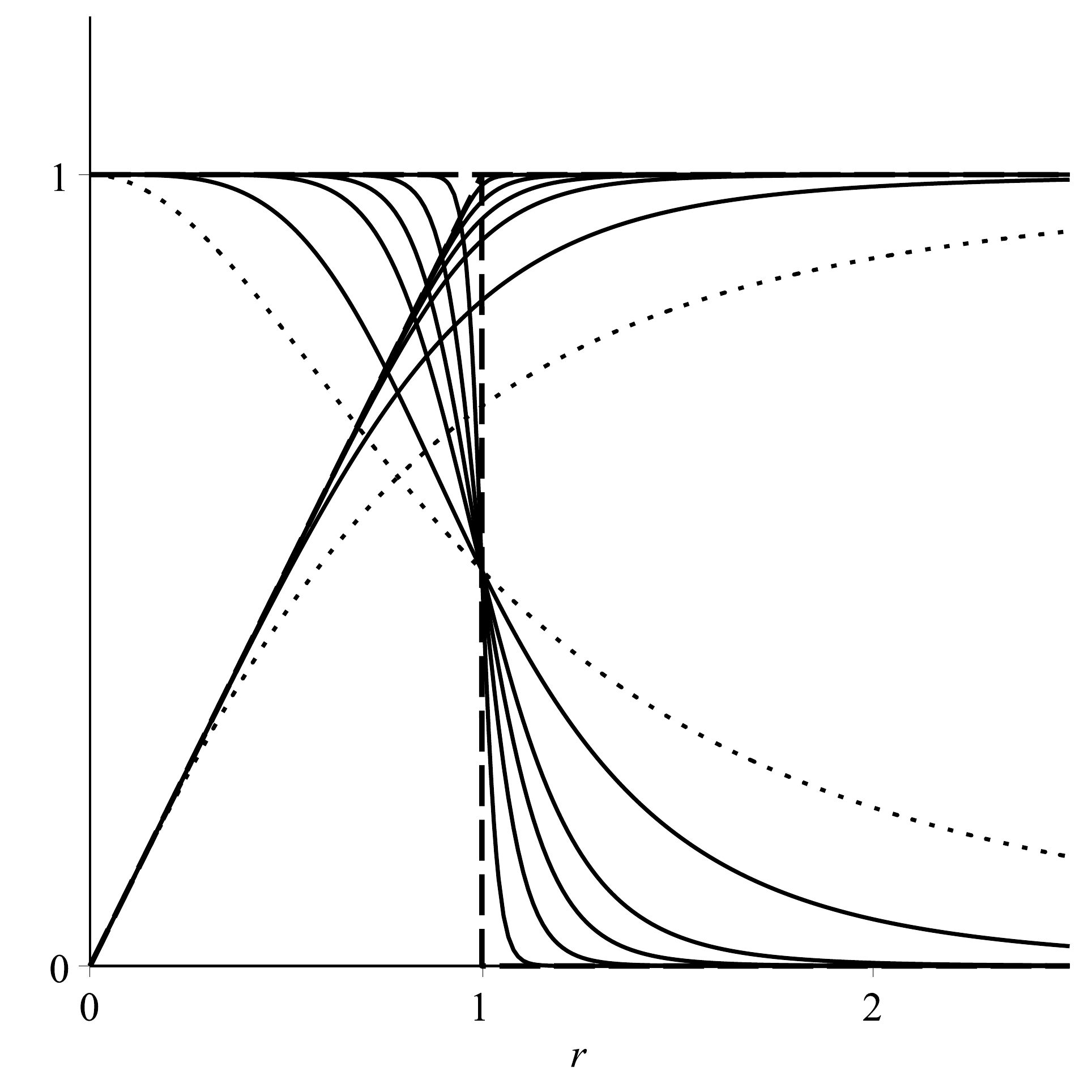}
\includegraphics[width=4.2cm]{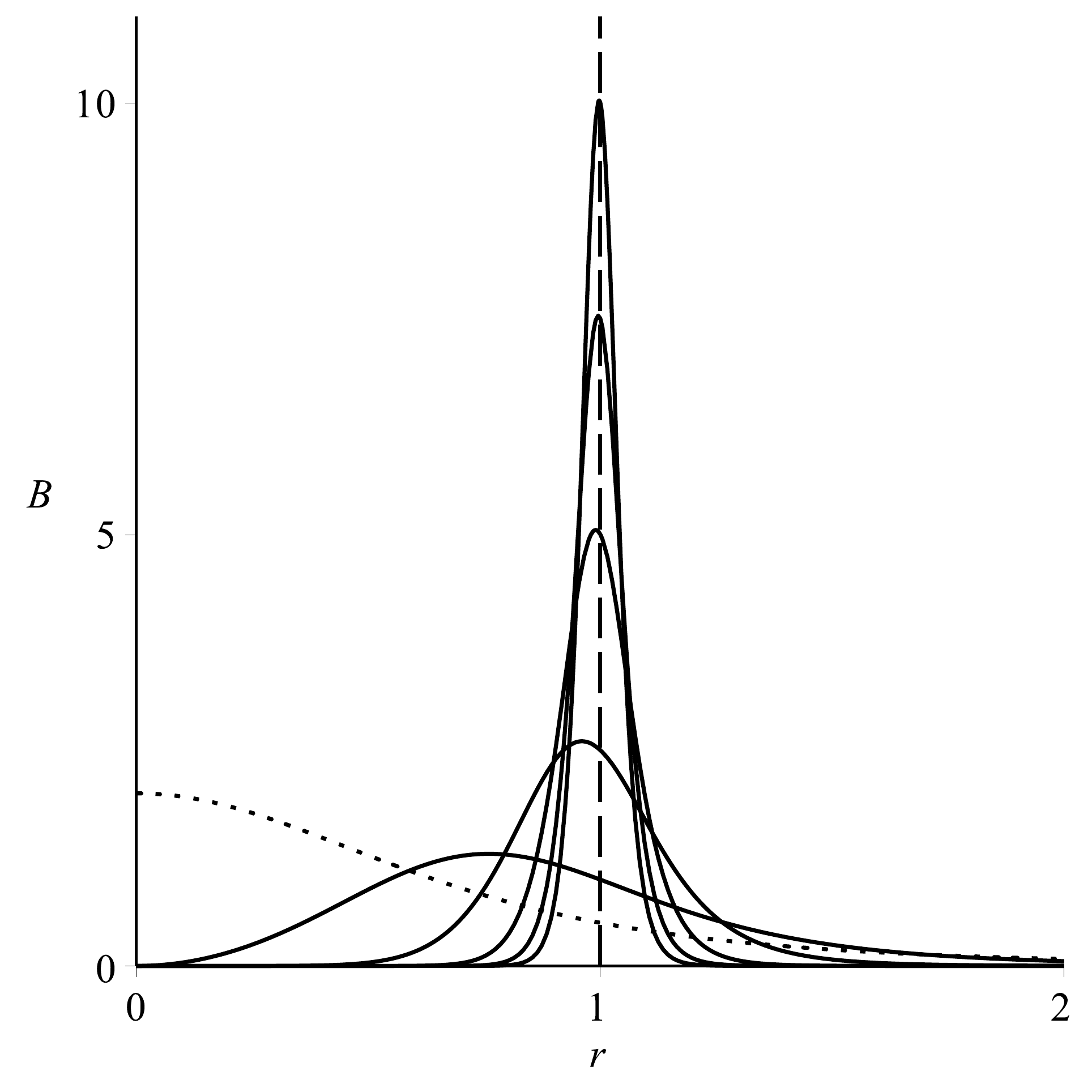}
\caption{In the left panel, we display the solutions $a(r)$ (descending lines) and $g(r)$ (ascending lines) given by Eqs.~\eqref{ag2}. In the right panel one depicts the magnetic field obtained in Eq.~\eqref{BR2}. The dotted lines represent the case $l=1$ and the dashed lines stand for the compact limit.}
\label{figsolb2}
\end{figure}

The function $R(g)$ considered in Eq.~\eqref{R1} for the previous example can be generalized to
\be\label{R2}
R(g)=g\left(1-g^{2l}\right),
\ee
where $l$ is a real parameter such that $l\geq1$. By substituting the above function in Eqs.~\eqref{dg1} and \eqref{arg}, one gets the solutions
\be\label{ag2}
g(r) = \frac{r}{\left(1+r^{2l}\right)^{\frac{1}{2l}}}, \quad\text{and}\quad a(r) =\frac{1}{1+r^{2l}}.
\ee
In this case, the magnetic field is
\be\label{BR2}
B(r)=\frac{2lr^{2l-2}}{(1+r^{2l})^2}.
\ee
As in the previous example, we can see that both Eqs.~\eqref{ag2} and \eqref{BR2} only depends on the function $R$ given by Eq.~\eqref{R2}. Furthermore, one can show that, as $l$ increases to larger and larger values, the solutions in Eqs.~\eqref{ag2} tend to assume the compact profile
\bes\label{solc}
\bal
a_c(r) &=
\begin{cases}
1,\,\,&r\leq 1\\
0, \,\, & r>1,
\end{cases}  \\
g_c(r) &=
\begin{cases}
r,\,\,\,&r\leq 1\\
1, \,\,\, & r>1,
\end{cases}
\eal
\ees
The magnetic field in Eq.~\eqref{BR2} also presents a compact limit, given by
\be\label{bc}
B_c(r) = \frac{\delta(r-1)}{r},
\ee
where $\delta(z)$ is the Dirac delta function. Although the solutions in Eq.~\eqref{ag2} were suggested in Ref.~\cite{ana1}, there the model was not constructed. Moreover, the compact limit was not studied for the solutions, magnetic field and energy density. We then go on and in Fig.~\eqref{figsolb2} one depicts the solutions \eqref{ag2} and the magnetic field \eqref{BR2}, for several values of $l$, including the compact limit in Eqs.~\eqref{solc} and \eqref{bc}.

\begin{figure}[htb!]
\centering
\includegraphics[width=5cm]{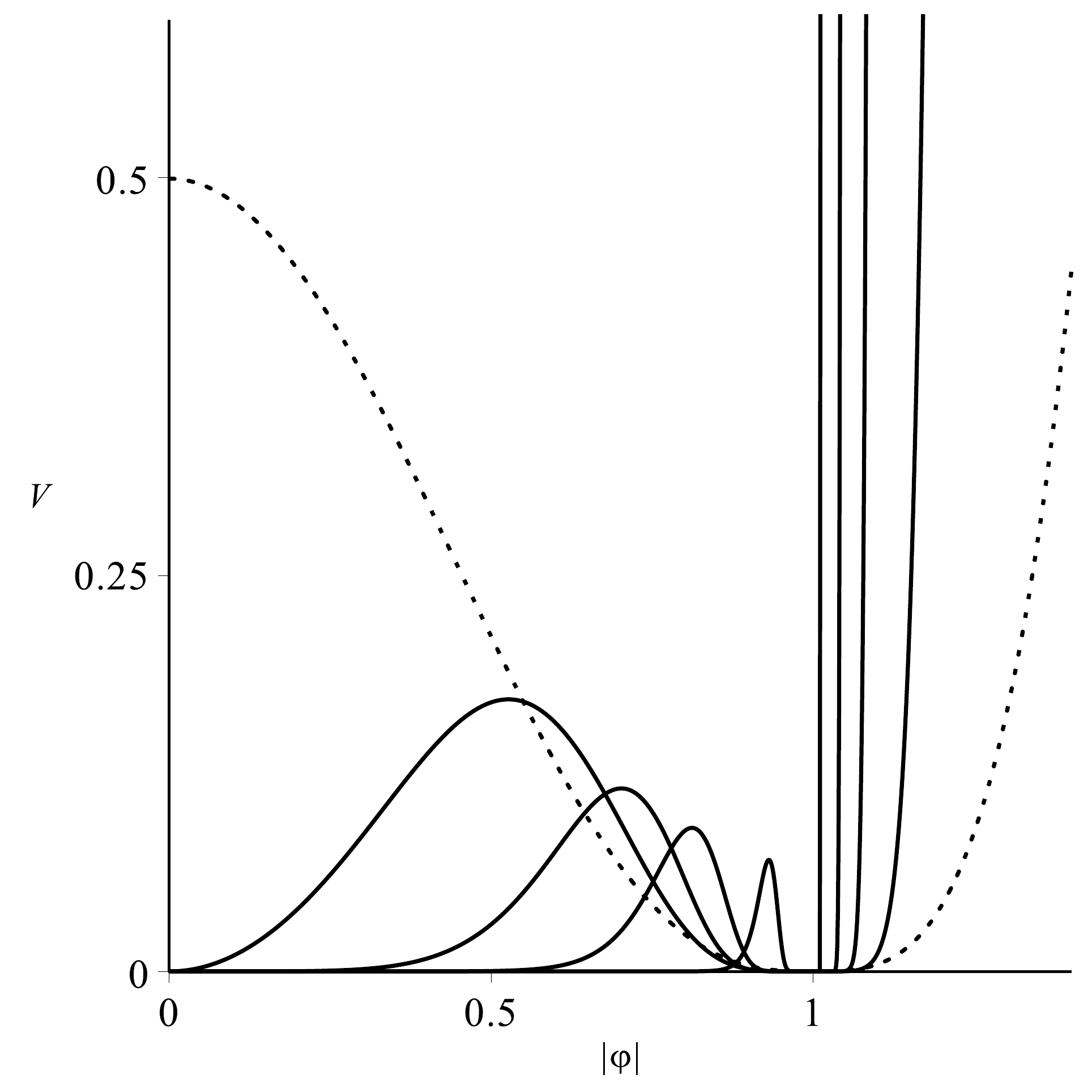}
\caption{The potential in Eq.~\eqref{pot2} is depicted for $\alpha=3$ and for $l=1,2,4,8$ and $32$. The dotted line represents the case $l=1$.}
\label{figpot2}
\end{figure}

The inverse function of $g(r)$ gives $q(g)$. Combining it with $R(g)$ in Eq.~\eqref{R2}, we can calculate $M(g)$ by using Eq.~\eqref{M}. Thus, we have
\bal
q(g) &= \frac{g}{\left(1-g^{2l}\right)^{\frac{1}{2l}}}, \\
M(g) &= -2lg^{2l-2} \left(1-g^{2l}\right)^{1+\frac{1}{l}}.
\eal
To find the functions $K(|\vphi|)$, $P(|\vphi|)$ and $V(|\vphi|)$, we must use Eqs.~\eqref{VKP}. As we discussed before, one can choose the potential
\be\label{pot2}
V(|\vphi|) = \frac{1}{2} l|\vphi|^{2l-2} \left|1-|\vphi|^{2l}\right|^{\alpha l},
\ee 
where $\alpha>2$. The minima of this potential are located at $|\vphi|=0$ and at $|\vphi|=1$ for any $l\neq1$. The local maximum is at $|\vphi_{max}| = \left((l-1)/(l(\alpha l+1)-1) \right)^{1/(2l)}$, such that the potential assumes the maximum value $V(|\vphi_{max}|) = l^{2\alpha l+1}(l-1)^{1-1/l}(l(\alpha l+1) -1)^{-\alpha l -1 + 1/l} \alpha^{\alpha l}/2$ inside the interval $|\varphi|\in[0,1]$. In the limit $l\to\infty$, we have
$|\vphi_{max}| \to 1$ and $V(|\vphi_{max}|) \to 1/(2e\alpha)$. In Fig.~\ref{figpot2}, we display the potential in Eq.~\eqref{pot2} for several values of $l$. The above choice for the potential leads to
\bal
K(|\vphi|) &= \frac12 \left(\alpha l^2 -l -1\right) |\vphi|^{2l-2} \left|1-|\vphi|^{2l}\right|^{\alpha l -2 -1/l},\\
P(|\vphi|) &= \frac{1}{4l} |\vphi|^{2-2l} \left|1-|\vphi|^{2l}\right|^{\alpha l -2 -2/l}.
\eal

Furthermore, one can use Eq.~\eqref{w} to calculate the function $W(a,g)$
\be\label{w2}
W(a,g) = -\frac12 a \left(1-g^{2l}\right)^{\alpha l -1 -1/l}.
\ee
From Eq.~\eqref{energyw}, it is straightforward to show that the energy is $E=\pi$.
To calculate the energy density, we use Eq.~\eqref{rhog}, which leads to
\be
\rho(g(r))= \left(\alpha l^2 -1\right)g^{2l-2} \left(1-g^{2l}\right)^{\alpha l}.
\ee
By using the solutions \eqref{ag2}, the explicit form of the energy density is
\be\label{rho2}
\rho(r) = \frac{\left(\alpha l^2 -1\right)r^{2l-2}}{\left(1 + r^{2l}\right)^{\alpha l +1 -1/l}}.
\ee
One can integrate the above expression to get the energy $E=\pi$, which is the same obtained by using the auxiliary function $W(a,g)$ as in Eq.~\eqref{w2}.
\begin{figure}[htb!]
\centering
\includegraphics[width=5cm]{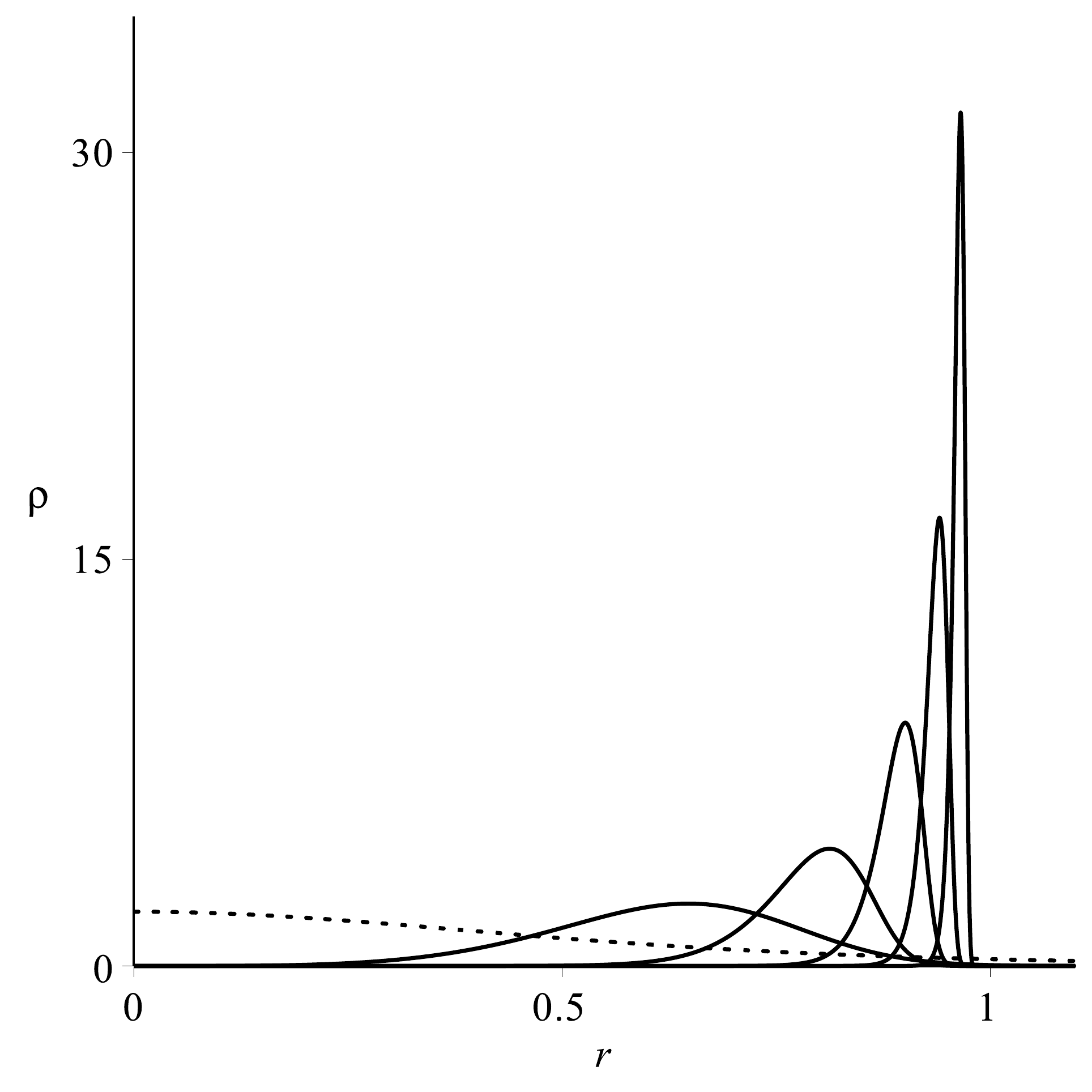}
\caption{The energy density in Eq.~\eqref{rho2} is displayed for $\alpha=3$ and for several values of $l$. The dotted line represent the case $l=1$.}
\label{figrho2}
\end{figure} 
We see from Fig.~\ref{figrho2} that the energy density tend to compactify near $r=1$. We can show this behavior analytically by taking into account that, for a general $l$, the peak in the energy density is at 
\be
r_p=\left(\frac{l-1}{\alpha l^2}\right)^{\frac{1}{2l}}.
\ee
By substituting this in Eq.~\eqref{rho2}, we get
\be
\rho(r_p) = \frac{(\alpha l^2)^{\alpha l} (\alpha l^2 - 1) (l-1)^{1-1/l}}{ ((\alpha l +1)\, l-1)^{\alpha l +1 -1/l}}.
\ee
For $l\to\infty$, we have $r_p \to 1$ and $\rho(r_p)\to \infty$. This means that, as $l$ increases, the peak of the energy density approaches $r=1$ and become taller and taller. Therefore, in the limit $l\to\infty$, energy of the vortex tend to become fully concentrated at $r=1$. Since the energy is $\pi$ for any $l$, we can conclude that the energy density in the compact limit has the form
\be
\rho_c(r) = \frac12\delta(r-1).
\ee
This behavior is similar to the one found in Ref.~\cite{compcs} for Chern-Simons-Higgs vortices, in which the energy density of the vortex tend to compactfy around a given radius, forming a ring in the polar plane. Nevertheless, we are dealing with Maxwell-Higgs vortices in this paper and we have the analytical expressions for a general $l$, which allow us to see how the compactification happens analytically, not numerically, as it was done in Ref.~\cite{compcs}. 

\begin{figure}[t]
\centering
\includegraphics[width=3cm]{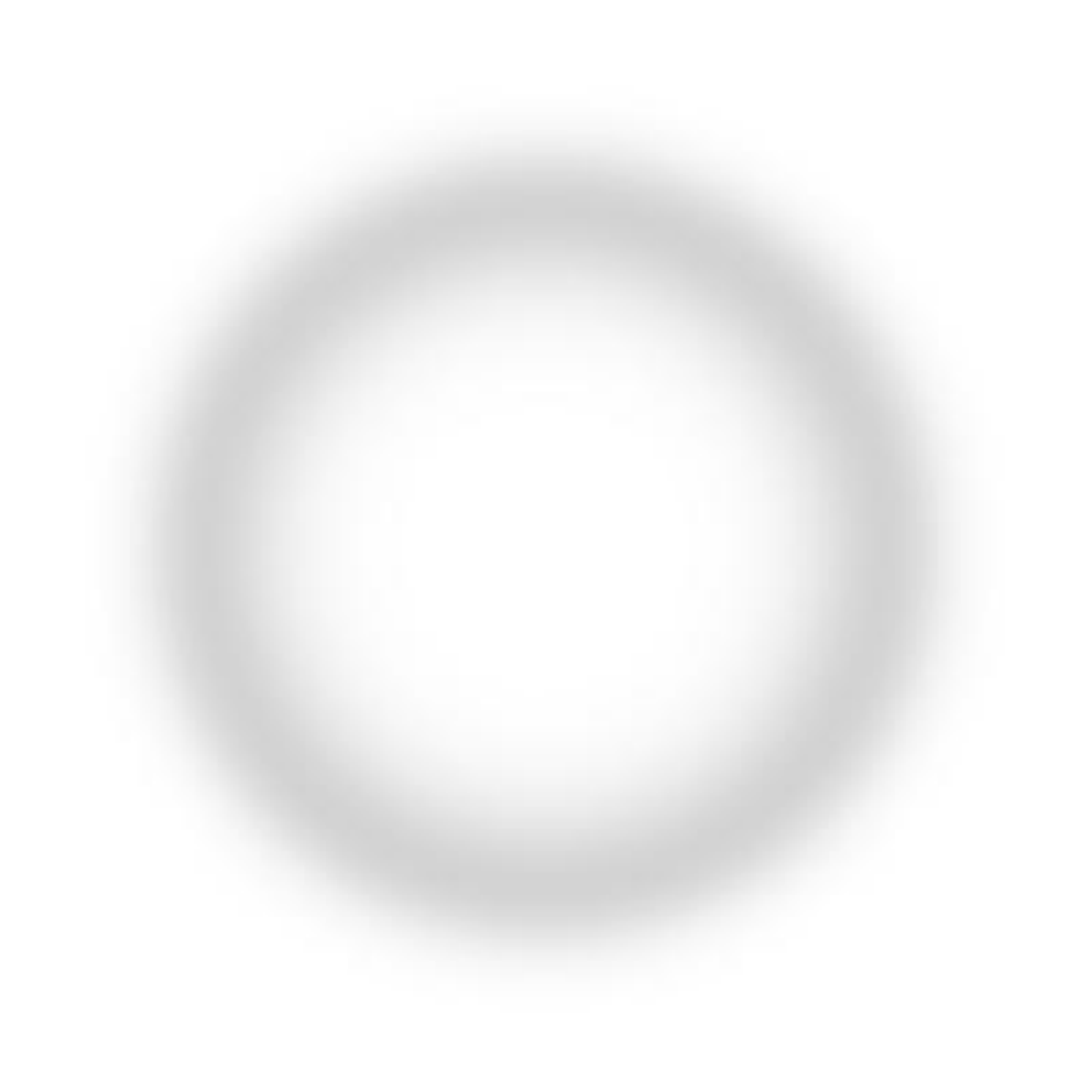}
\includegraphics[width=3cm]{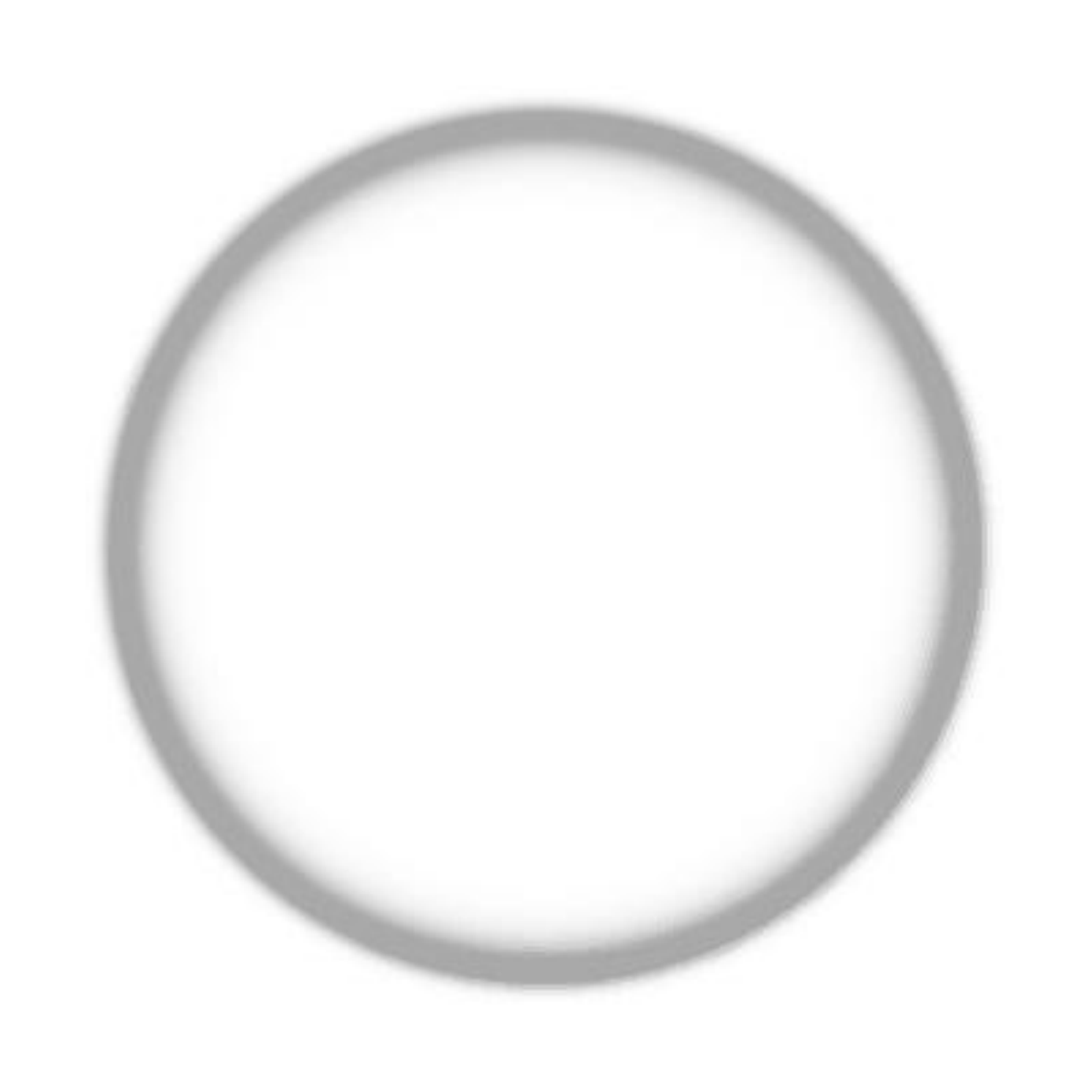}
\includegraphics[width=3cm]{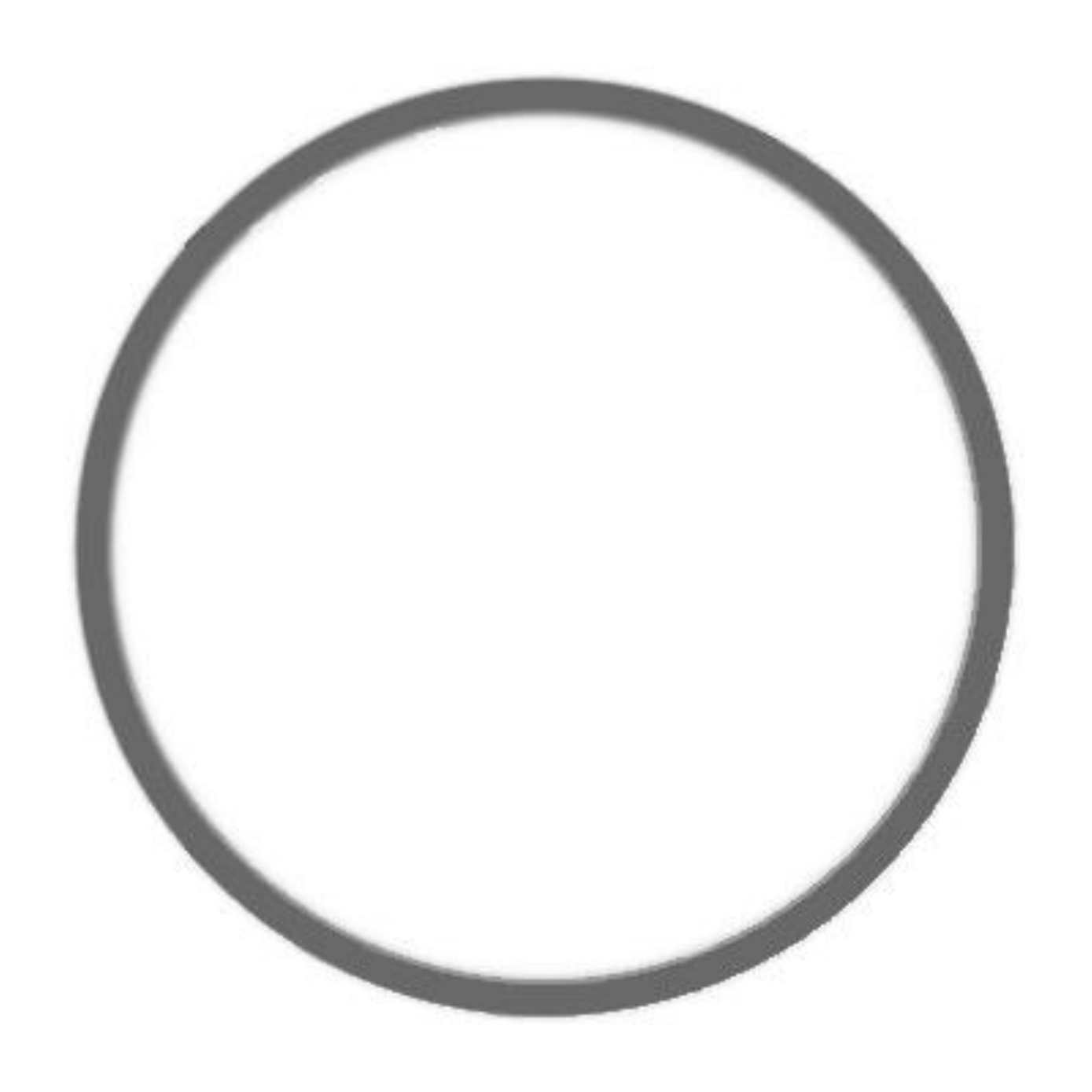}
\includegraphics[width=3cm]{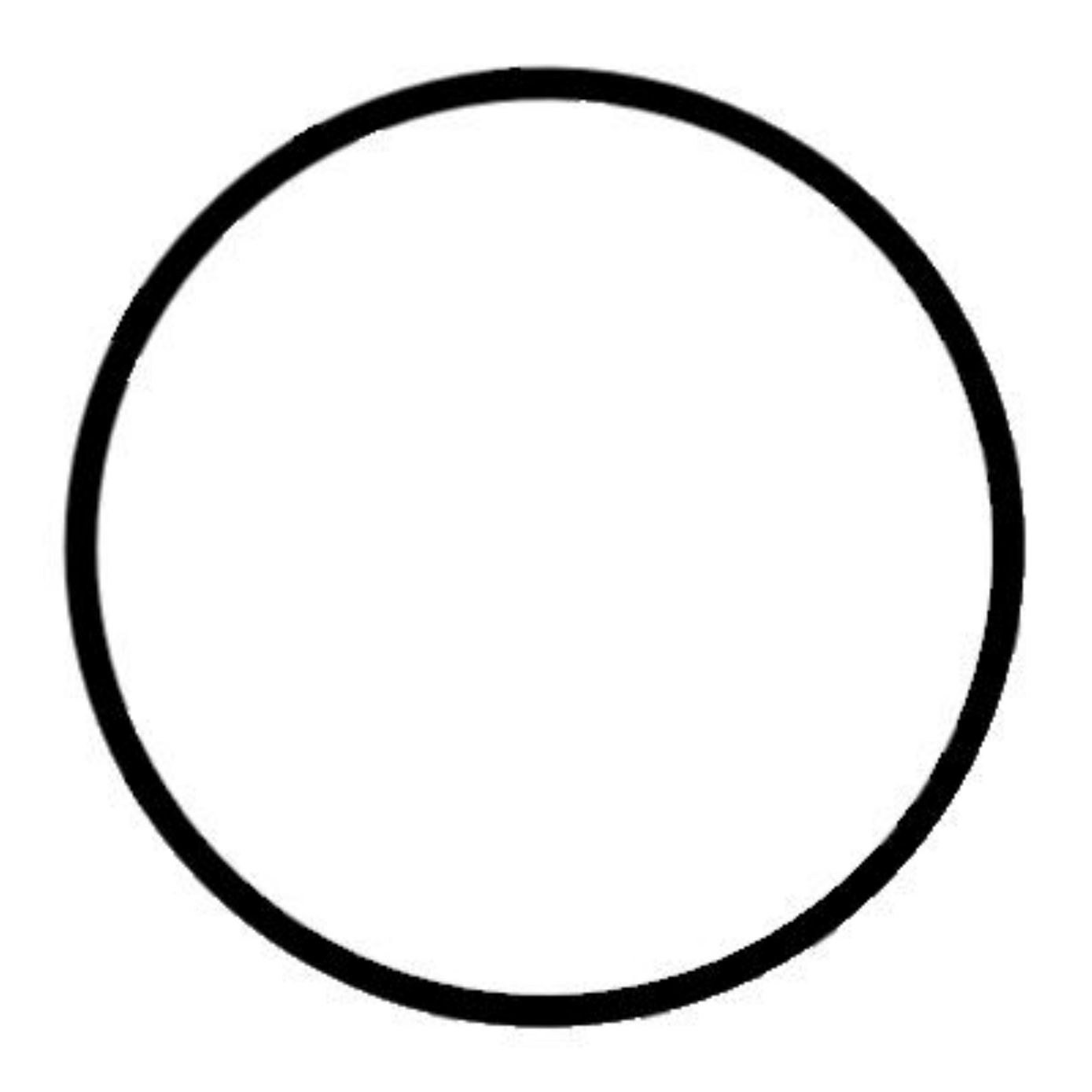}
\caption{The energy density in Eq.~\eqref{rho2}, depicted for $\alpha=3$ and for $l=4,16, 64$ and $256$, showing how the solution behaves as $l$ increases to the compact limit.}
\label{fig10}
\end{figure}

As far as we can see, the above behavior was never seen before in a Maxwell-Higgs model, so we display it again in Fig.~\ref{fig10}, in the $(r,\theta)$ plane, in order to better illustrate the behavior of the analytical vortex configuration found above, showing how the solution shrinks to the circle with $r_p\to1$ as $l$ increases to infinite.

\section{Ending comments}
\label{sec:4}

In this work we studied the presence of vortices in generalized models of the Maxwell-Higgs type.  The models that we consider contain two new functions of the scalar field, $P(|\varphi|)$ and $K(|\varphi|)$, which are used to control the Maxwell dynamics and the covariant derivative, respectively. Although this kind of generalization is not new, in the current work we have developed a procedure that allows the presence of first-order differential equations that solve the equations of motion under certain circumstances. 

The approach suggests a way to decouple the first-order equations as an important step to find analytic solutions. Also, it includes an auxiliary function, from which one can write the energy of the vortex exactly, calculated from the asymptotic values of the scalar and gauge fields. 
The method proposed to decouple the first-order equations requires the construction of the Maxwell-Higgs model, and here it is implemented by suggesting the explicit form of the potential, and then working to collect the two functions $P(|\varphi|)$ and $K(|\varphi|)$ and write the Lagrangian density explicitly. 

We illustrated the procedure with three distinct examples, which work adequately and suggest that the methodology is robust. In particular, we have studied a model that supports analytic configurations controlled by a parameter that may induce the presence of vortices with compact behavior. 

The results described in this work may stimulate new investigations on related issues, in particular  on the extension of the study to the case of the Chern-Simons, Dirac-Born-Infeld, Skyrme and other generalized dynamics, and also in the presence of non-Abelian fields, to search for topological configurations that can be expressed analytically. They may also be of interest to holography, as suggested, for instance, by \cite{H0,H1,H2} and references therein. Since the model contains generalized magnetic permeability, one may investigate the presence of vortices in an electromagnetic metamaterial, in a medium with negative refractive index \cite{N0,N1,N2}. 

In the case of the Dirac-Born-Infeld (DBI) modification, a recent study \cite{db} investigated several models that present BPS-like configurations, and they may also be of direct interest as generalized models to be extended to support vortex configurations as the ones investigated in the current work. Another line of investigation concerns General Relativity (GR): one knows that GR faces problems at high energies, near the Planck scale, and
this suggests the need of modifications; in this sense, the DBI inspired modifications of gravity that have been reviewed in \cite{olmo} and the current study may also suggest new investigations in the subject.

 Another possibility of current interest is to enlarge the model \eqref{lmodel} and consider, for instance, the $U(1)\times U(1)$ model
\ben\label{large}
\LL &=&- \frac{1}{4}P_v(|\vphi|,|\chi|)F_{\mu\nu}F^{\mu\nu}+ K_v(|\vphi|,|\chi|)|D_{\mu}\vphi|^2\nonumber\\
&&- \frac{1}{4}P_h(|\vphi|,|\chi|){\cal F}_{\mu\nu}{\cal F}^{\mu\nu}+ K_h(|\vphi|,|\chi|)|{\cal D}_{\mu}\chi|^2 \nonumber\\
&& -\lambda F_{\mu\nu}{\cal F}^{\mu\nu}- V(|\vphi|,|\chi|),
\een
where ${\cal F}_{\mu\nu}=\partial_\mu {\cal A}_\nu-\partial_\nu {\cal A}_\mu$ and ${\cal D}_\mu=\partial_\mu-iq {\cal A}_\mu$, with $\chi$ and ${\cal A}_\mu$ being the second scalar and gauge fields, respectively, belonging to the hidden sector, in a way similar to the cases investigated in \cite{DM4,DM5}. This generalized model allows for several distinct possibilities of coupling between the visible and hidden sectors, and the presence of analytical solutions as the ones investigated in the current work may perhaps help us identify the correct coupling. The model \eqref{large} is under investigation and we hope to report on it in the near future.

\acknowledgements{ We would like to acknowledge the Brazilian agency CNPq for financial support. DB thanks support from fundings 455931/2014-3 and 306614/2014-6, LL thanks support from fundings 307111/2013-0 and 447643/2014-2, MAM thanks support from funding 140735/2015-1 and RM thanks support from fundings 455619/2014-0 and 306826/2015-1.}

\end{document}